\documentclass[
 aip,
 apl,
 amsmath,amssymb,
 reprint,
 floatfix,
 article-title,
 superscriptaddress
]{revtex4-1}
\usepackage{hyperref}
\usepackage{graphicx}
\usepackage{dcolumn}
\usepackage{bm}
\usepackage[utf8]{inputenc}
\usepackage[T1]{fontenc}
\usepackage{mathptmx}
\usepackage{etoolbox}
\usepackage[UTF8,scheme=plain]{ctex}
\usepackage{hyperref}
\hypersetup{hidelinks}

\makeatletter
\def\@email#1#2{%
 \endgroup
 \patchcmd{\titleblock@produce}
  {\frontmatter@RRAPformat}
  {\frontmatter@RRAPformat{\produce@RRAP{*#1\href{mailto:#2}{#2}}}\frontmatter@RRAPformat}
  {}{}
}%
\makeatother

\begin{document}

\title{Lensless single-pixel imaging through dynamic coded-aperture illumination}

\author{Yuhan Guo}
\thanks{\ These authors contributed equally to this work.}
\author{Shilei Li}
\thanks{\ These authors contributed equally to this work.}
\author{Tong Zhou}
\author{Liming Li\textsuperscript{*}}
\email{liliming@sdut.edu.cn}
\date{\today}

\begin{abstract}
We demonstrate a lensless single-pixel imaging (SPI) system combining dynamic coded-aperture illumination with single-pixel detection. Binary digital micromirror device (DMD) masks generate tailored grayscale intensity fields through coherent propagation, without relay projection optics or temporal grayscale synthesis. Six target-field configurations---uniform, Rayleigh speckle, squared-Rayleigh speckle, and sparse binary fields with $p=10\%$, $5\%$, and $1\%$---are evaluated using binary and grayscale objects. Intensity-correlation and nonnegative total-variation-regularized reconstructions reveal that the preferred illumination depends on the object, reconstruction method, and evaluation metric. By combining rapid binary modulation, programmable grayscale illumination, and single-pixel readout, this implementation provides a flexible architecture for lensless computational imaging and joint design of optical encoding and reconstruction.
\end{abstract}

\maketitle

Pinhole imaging offers lensless operation but involves a trade-off between photon throughput and spatial resolution. Coded-aperture imaging uses multi-aperture masks and computational decoding to collect and separate spatially multiplexed signals~\cite{Fenimore80}. Its structural simplicity has motivated applications in computational imaging, including high-energy imaging~\cite{braga2020coded}. Combining dynamic aperture encoding with single-pixel detection offers a complementary route where spatial information is recovered from sequential scalar measurements rather than detector-array images.

Single-pixel imaging (SPI) reconstructs an object from known spatial modulations and the corresponding bucket signals\cite{shapiro2008computational,bromberg2009ghost,edgar2019principles}. This approach is attractive when array detectors are costly or inefficient\cite{wang2023midinfrared}. Applications include three-dimensional imaging \cite{sun2013three}, underwater imaging \cite{Yang21}, and X-ray imaging \cite{pelliccia2016xray,yu2016fourier}. In an illumination-side coded-aperture implementation, coherent diffraction generates the spatial encodings at the object. Sequential aperture modulation and single-pixel readout therefore provide a high-speed architecture without requiring a spatially resolving detector.

Digital micromirror devices (DMDs) offer rapid binary modulation\cite{duarte2008single,goorden2014superpixel,hahamovich2021megahertz,kilcullen2022compressed}. Direct projection of their masks typically requires relay optics and produces binary illumination, whereas computer-generated holography (CGH) enables grayscale fields from binary amplitude masks\cite{lee1979binary,hoyos2024non}. In our previous study, we proposed a holographic ghost-imaging scheme using a phase-only spatial light modulator and camera-derived bucket signals\cite{li2026integrating}.

Here, we experimentally implement this architecture by integrating a DMD, a single-pixel detector, and a digital acquisition card. Gerchberg–Saxton (GS) optimization and Otsu thresholding\cite{gerchberg1972practical,fienup1982phase,otsu1979threshold} encode each illumination into one binary aperture pattern. Coherent propagation produces grayscale fields without relay projection optics, while numerical propagation of the binary aperture pattern provides the reconstruction reference signals. Six target-field configurations are evaluated for binary and grayscale objects using intensity correlation and nonnegative total-variation (TV) regularization. The comparisons establish how programmable aperture encoding interacts with object structure, reconstruction, and image-quality metrics.

Customized projection via coherent multi-aperture illumination follows a three-step procedure. First, the complex diffraction field ${E}_{\mathrm{GS}}$ corresponding to the target intensity $I_{\mathrm{T}}$ is retrieved with the GS algorithm. Second, a quadratic propagation phase is added, and the real part $\Re\{\cdot\}$ of the resulting field is retained. Finally, this grayscale pattern is binarized by Otsu thresholding. The grayscale hologram precursor is
\begin{equation}\label{eq:aogr}
\begin{split}
AO_{\mathrm{GR}}(r)
 &= \Re\!\left\{{E}_{\mathrm{GS}}\exp\!\left(-{ikr^2/(2z_0)}\right)\right\},
\end{split}
\end{equation}
where $r$ is the transverse coordinate in the CGH plane, $k=2\pi/\lambda$ is the wave number, and $z_0$ is the DMD-to-object propagation distance. Experimentally, restricting the target support to one half of the reconstruction plane spatially separates the desired order from its conjugate.

The binary-amplitude CGH is then given by
\begin{equation}\label{eq:aobi}
AO_{\mathrm{BI}}(r)=
\begin{cases}
1, & AO_{\mathrm{GR}}(r)\ge t^{*},\\
0, & AO_{\mathrm{GR}}(r)<t^{*},
\end{cases}
\end{equation}
where $t^{*}$ is the threshold selected by Otsu's criterion~\cite{otsu1979threshold}. Thus, $AO_{\mathrm{BI}}$ acts as a coded multi-aperture mask whose coherent diffraction approximates the prescribed grayscale field. Loading a sequence of these masks on the DMD provides the time-varying illumination required for lensless SPI.

For the $m$th projected field $I_m(r)$, the bucket detector records
\begin{equation}\label{eq:bucket}
b_m=\int_{\Omega}T(r)I_m(r)\,\mathrm{d}r+n_m,
\end{equation}
where $T(r)$ is the object intensity transmittance, $\Omega$ is the illuminated region, and $n_m$ represents detection and environmental noise.
For all reconstructions reported here, the computational reference $I_m(r)$ is the object-plane intensity obtained by numerically propagating the $m$th displayed DMD hologram under the experimental geometry. The conventional correlation image is evaluated as
\begin{equation}\label{eq:correlation}
G(r)=\left\langle b_m I_m(r)\right\rangle_m
-\left\langle b_m\right\rangle_m\left\langle I_m(r)\right\rangle_m.
\end{equation}

Compressive sensing can reduce the required number of measurements $M$ by exploiting object priors\cite{duarte2008single,katz2009compressive}. Let $\mathrm{S}\in\mathbb{R}^{M\times P}$ collect the vectorized object-plane intensity patterns (one pattern per row), and let $\mathrm{b}_0\in\mathbb{R}^{M}$ contain the corresponding bucket signals. Here, $P$ is the number of pixels in each pattern. We adopt a covariance-form model: $\mathrm{S}$ and $\mathrm{b}_0$ are mean-centered and normalized by the root-mean-square value $\mathrm{R}_\mathrm{S}$ and standard deviation $\mathrm{S}_{\mathrm{b}_0}$, respectively, to yield
\begin{equation}\label{eq:normalized_model}
\widetilde{\mathbf{A}}=
{(\mathrm{S}-\mathbf{1}\overline{\mathbf{s}}^{\mathsf T})}/{(\mathrm{R}_\mathrm{S}\sqrt{M})},
\quad
\widetilde{\mathrm{b}}=
{\eta(\mathrm{b}_0-\overline{\mathrm{b}}_0\mathbf{1})}/{(\mathrm{S}_{\mathrm{b}_0}\sqrt{M})},
\end{equation}
where $\overline{\mathbf{s}}$ is the mean pattern across measurements, $\overline{\mathrm{b}}_0$ is the mean bucket signal, and $\mathbf{1}$ is an $M$-element vector of ones. The factor $\eta\in\{-1,1\}$ corrects the detection polarity. The reconstructed image $\mathbf{X}$ is obtained by solving the nonnegative TV-regularized problem:
\begin{equation}\label{eq:tv}
\widehat{\mathbf{X}}=
\arg\min_{\mathbf{X}\ge0}
\left\{
\frac{1}{2}\left\|\widetilde{\mathbf{A}}\operatorname{vec}(\mathbf{X})-
\widetilde{\mathrm{b}}\right\|_2^2
+\lambda_{\mathrm{TV}}\operatorname{TV}(\mathbf{X})
\right\},
\end{equation}
with isotropic total variation
\begin{equation}\label{eq:isotropic_tv}
\operatorname{TV}(\mathbf{X})=
\sum_{i,j}\sqrt{(D_x\mathbf{X})_{i,j}^{2}+(D_y\mathbf{X})_{i,j}^{2}}
\end{equation}
serving as a piecewise-smooth image prior\cite{rudin1992nonlinear}. Equation~\eqref{eq:tv} is solved using the Chambolle--Pock primal--dual algorithm\cite{chambolle2011first}, with $\lambda_{\mathrm{TV}}=0.05$ for all reconstructions. Reconstruction quality was evaluated using peak signal-to-noise ratio (PSNR), contrast-to-noise ratio (CNR), and structural similarity index (SSIM). Details of the metric definitions can be found in the supplementary material.

\begin{figure}[b]
\centering
\includegraphics[width=0.95\columnwidth]{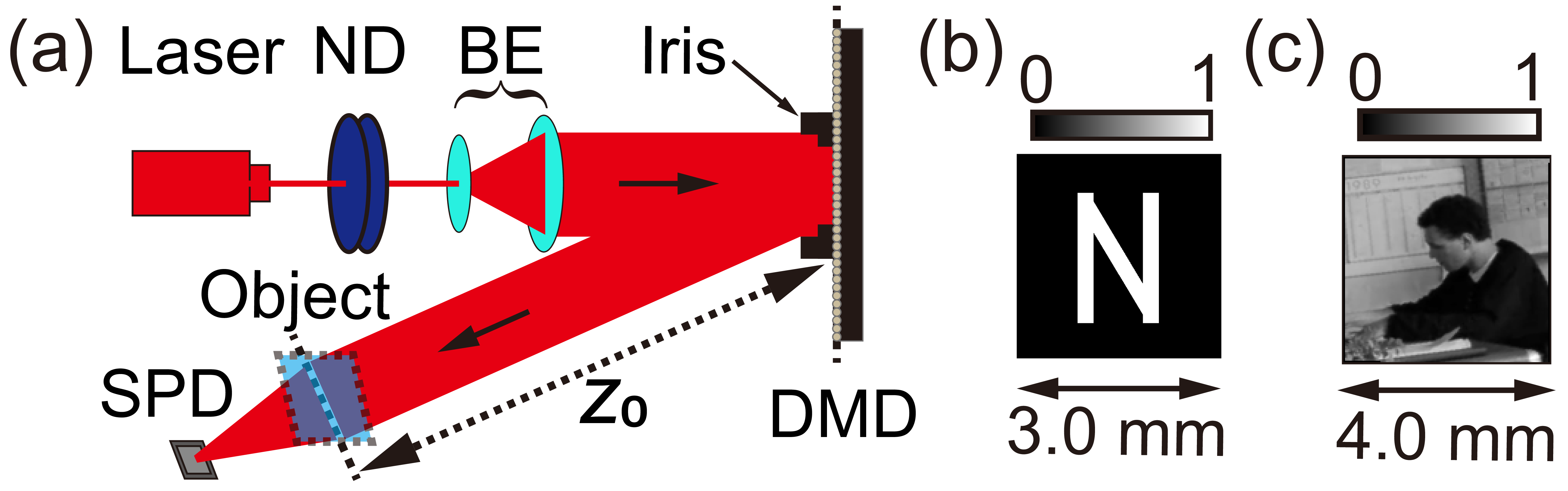}
\caption{(a) Lensless SPI setup using coherent dynamic multi-aperture illumination. ND, neutral-density filter; BE, beam expander; DMD, digital micromirror device; SPD, single-pixel detector. Two test objects: (b) a binary “N” and (c) a grayscale portrait.}\label{fig01:setup}
\end{figure}

Figure~\ref{fig01:setup} shows the experimental arrangement and the two test objects. A single-mode continuous-wave laser at $\lambda=632.8$~nm was attenuated and expanded before illuminating a DMD (DLP6500FYE, Texas Instruments) with a pixel pitch of $7.56~\mu\mathrm{m}$ and a resolution of $1920\times1080$ pixels. An iris positioned close to the DMD defined an active hologram aperture of diameter $D=4.0$~mm. The modulated field propagated by free-space diffraction over $z_0=0.6$~m to the object plane. A single-pixel detector (PDA100A2, Thorlabs) recorded the transmitted signal. The detector output was digitized at $125\,\mathrm{MSPS}$ using a digital acquisition card (DAQ207, Lockzhiner Electronic) with an AD9268 analog-to-digital converter (Analog Devices). Operating at 10 kHz, the DMD enabled the acquisition of 10,000 measurements in one second. No projection or imaging lens was used between the DMD, object, and detector. The binary and grayscale objects shown in Figs.~\ref{fig01:setup}(b) and \ref{fig01:setup}(c) were square, with side lengths of 3.0~mm and 4.0~mm, respectively.

To evaluate customized projection with dynamic multi-aperture illumination, we selected two categories of target fields: continuous-valued fields and sparse binary fields. The continuous-valued fields included a uniform field, a Rayleigh speckle field, and a squared-Rayleigh speckle field. The sparse binary fields were generated by randomly replacing selected elements of an all-zero matrix with ones. The nonzero pixels accounted for $p=10\%$, $5\%$, or $1\%$ of all pixels. All reconstructed images were normalized to $[0,1]$ for display and converted to 8-bit grayscale for fidelity evaluation. The corresponding computational reference fields were obtained by numerically propagating the target-field CGHs. Because the same procedure was applied to all six fields, only the sparse $p=1\%$ example is presented in the supplementary material.

\begin{figure}[b]
\centering
\includegraphics[width=0.9\columnwidth]{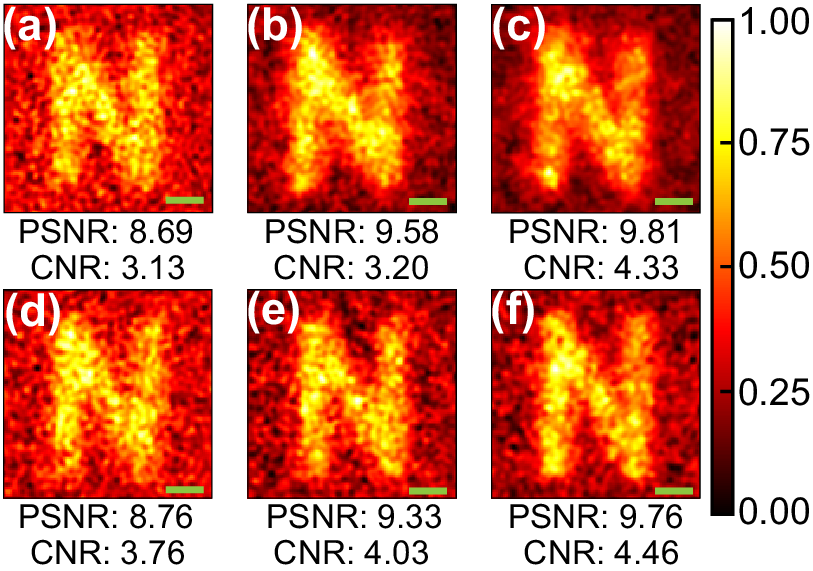}
\caption{Intensity-correlation reconstructions of the binary object ``N'' from $M=3,000$ measurements for (a)~a uniform field, (b)~Rayleigh speckle, (c)~squared-Rayleigh speckle, and sparse binary fields with (d)~$p=10\%$, (e)~$p=5\%$, and (f)~$p=1\%$. All panels share a common color scale. Scale bars: \(0.5~\mathrm{mm}\) in all panels.}\label{fig02:Ncorr}
\end{figure}

Using $M=3000$ measurements, Fig.~\ref{fig02:Ncorr} compares intensity-correlation reconstructions of the binary letter ``N'' under holographic illumination for six target fields. Because coherent holographic projection is affected by speckle noise, even the uniform target field produced a projected intensity distribution with Rayleigh-speckle characteristics. Replacing the uniform target with Rayleigh speckle and higher-dynamic-range squared-Rayleigh speckle\cite{bromberg2014nonrayleigh,zhou2017superbunching,wu2024phase} progressively increased the projected intensity contrast, accompanied by improved reconstruction quality. Accordingly, the PSNR increased from \(8.69\) dB for the uniform field [Fig.~\ref{fig02:Ncorr}(a)] to \(9.58\) dB for the Rayleigh field [Fig.~\ref{fig02:Ncorr}(b)] and \(9.81\) dB for the squared-Rayleigh field [Fig.~\ref{fig02:Ncorr}(c)]. For the sparse binary fields in Figs.~\ref{fig02:Ncorr}(d)--\ref{fig02:Ncorr}(f), decreasing $p$ from $10\%$ to $5\%$ and $1\%$ increased the PSNR from $8.76$\,dB to $9.33$\,dB and $9.76$\,dB, respectively. CNR was also evaluated to assess object--background contrast. The CNR increased from $3.13$ in Fig.~\ref{fig02:Ncorr}(a) to $3.20$ and $4.33$ in Figs.~\ref{fig02:Ncorr}(b) and~\ref{fig02:Ncorr}(c), respectively. Similarly, for the sparse binary fields in Figs.~\ref{fig02:Ncorr}(d)--\ref{fig02:Ncorr}(f), decreasing $p$ from $10\%$ to $5\%$ and $1\%$ increased the CNR from $3.76$ to $4.03$ and $4.46$, respectively. Nevertheless, the greater contrast of the sparse binary target fields did not yield a commensurate improvement in reconstruction quality compared with the continuous-valued target fields.

\begin{figure}[tbp]
\centering
\includegraphics[width=0.9\columnwidth]{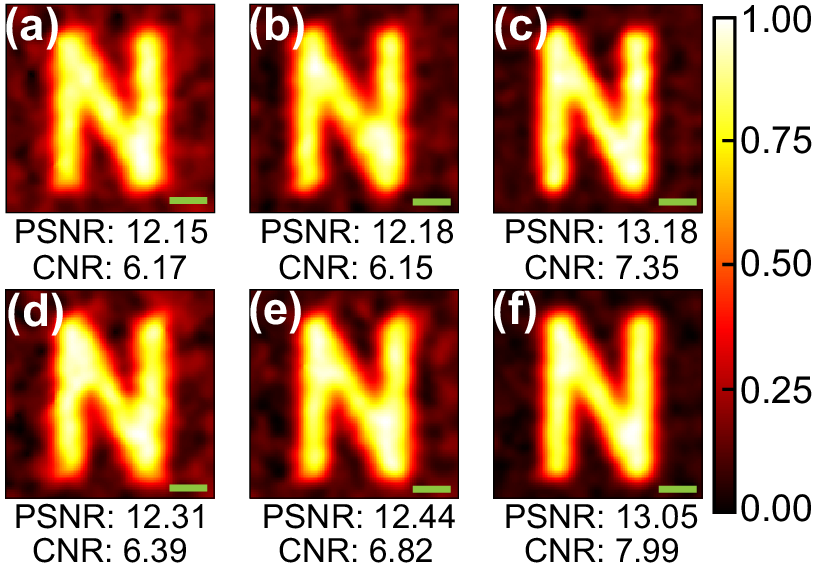}
\caption{Nonnegative TV-regularized reconstructions of the binary object under the same measurement conditions and target-field configurations as in Fig.~\ref{fig02:Ncorr}. All panels share the same color scale. Scale bars: \(0.5~\mathrm{mm}\).}\label{fig03:Ntv}
\end{figure}

The same measurement data used in Fig.~\ref{fig02:Ncorr} were reconstructed using nonnegative TV regularization, with the results shown in Fig.~\ref{fig03:Ntv}. Relative to intensity-correlation reconstruction, TV regularization visibly suppressed background fluctuations and increased the PSNR by $2.6$--$3.6$\,dB and the CNR by $2.6$--$3.5$ across all six target fields. Although the metrics did not increase strictly monotonically with target-field contrast in every case, the squared-Rayleigh speckle and the $p=1\%$ sparse binary field yielded the highest PSNR and CNR within their respective groups. 

\begin{figure}[b]
	\centering
	\includegraphics[width=0.9\columnwidth]{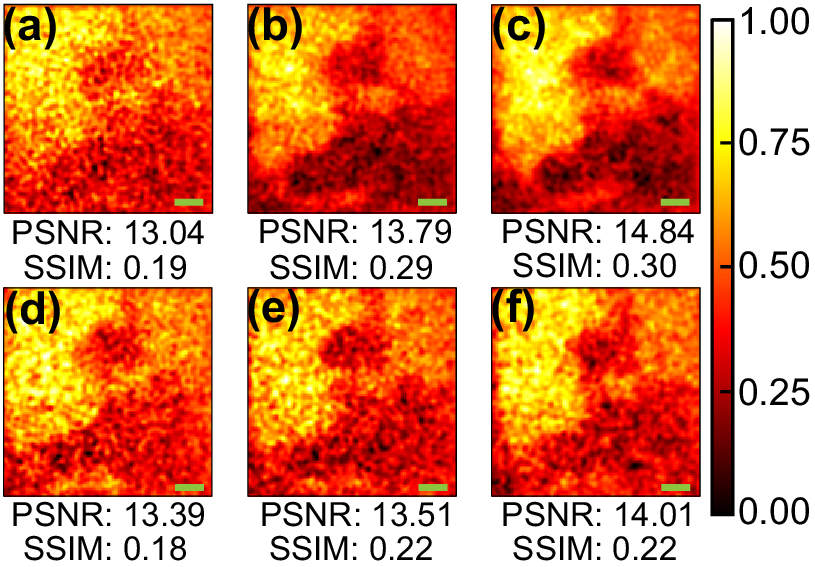}
	\caption{Intensity-correlation reconstructions of the grayscale object obtained under the same measurement conditions and target-field configurations as in Fig.~\ref{fig02:Ncorr}. All panels share a common color scale. Scale bars: \(0.5\) mm.}\label{fig04:Gcorr}
\end{figure}

\begin{figure}[t]
	\centering
	\includegraphics[width=0.9\columnwidth]{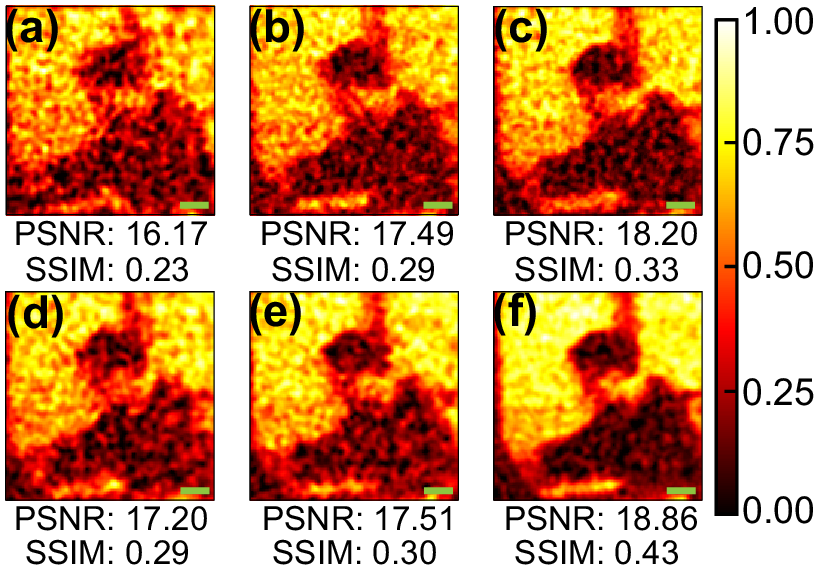}
	\caption{Nonnegative TV-regularized reconstructions of the grayscale object obtained from the same measurements used for the intensity-correlation reconstructions in Fig.~\ref{fig04:Gcorr}. All panels share a common color scale. Scale bars: \(0.5\) mm.}\label{fig05:Gtv}
\end{figure}
Figure~\ref{fig04:Gcorr} shows grayscale-portrait reconstructions at $M=3000$ using the same six target fields. For the continuous-valued target fields in Figs.~\ref{fig04:Gcorr}(a)--\ref{fig04:Gcorr}(c), the PSNR increased from $13.04$\,dB for the uniform field to $13.79$\,dB and $14.84$\,dB for the Rayleigh and squared-Rayleigh speckle fields, respectively. For the sparse binary fields in Figs.~\ref{fig04:Gcorr}(d)--\ref{fig04:Gcorr}(f), decreasing $p$ from $10\%$ to $5\%$ and $1\%$ increased the PSNR from $13.39$\,dB to $13.51$\,dB and $14.01$\,dB, respectively. Fine portrait details remained poorly resolved, motivating SSIM evaluation of structural fidelity. For the continuous-valued fields, the SSIM increased from $0.19$ in Fig.~\ref{fig04:Gcorr}(a) to $0.29$ and $0.30$ in Figs.~\ref{fig04:Gcorr}(b) and~\ref{fig04:Gcorr}(c), respectively. For the sparse binary fields, it increased from $0.18$ to $0.22$ as $p$ decreased from $10\%$ to $5\%$ and remained at $0.22$ for $p=1\%$. Overall, higher target-field contrast generally improved reconstruction quality, although the performance also depended on the target-field statistics and object structure.

We then applied nonnegative TV-regularized reconstruction to the same measurement data, with the results shown in Fig.~\ref{fig05:Gtv}. Relative to intensity-correlation reconstruction, TV regularization visibly suppressed background noise and improved the recovery of image details. Across all six target fields, the PSNR increased by $3.1$--$4.9$\,dB. The SSIM increased by $0.03$--$0.21$ for five fields but remained at $0.29$ for Rayleigh speckle at the reported precision. Reconstruction quality generally improved with increasing target-field contrast for the continuous-valued fields and with decreasing $p$ for the sparse binary fields. However, the best-performing target field depended on the reconstruction algorithm. The squared-Rayleigh speckle field yielded the highest PSNR and SSIM under intensity-correlation reconstruction, whereas the $p=1\%$ sparse binary field yielded the highest values under nonnegative TV-regularized reconstruction.

\begin{figure}[tbp]
	\centering
	\includegraphics[width=0.95\columnwidth]{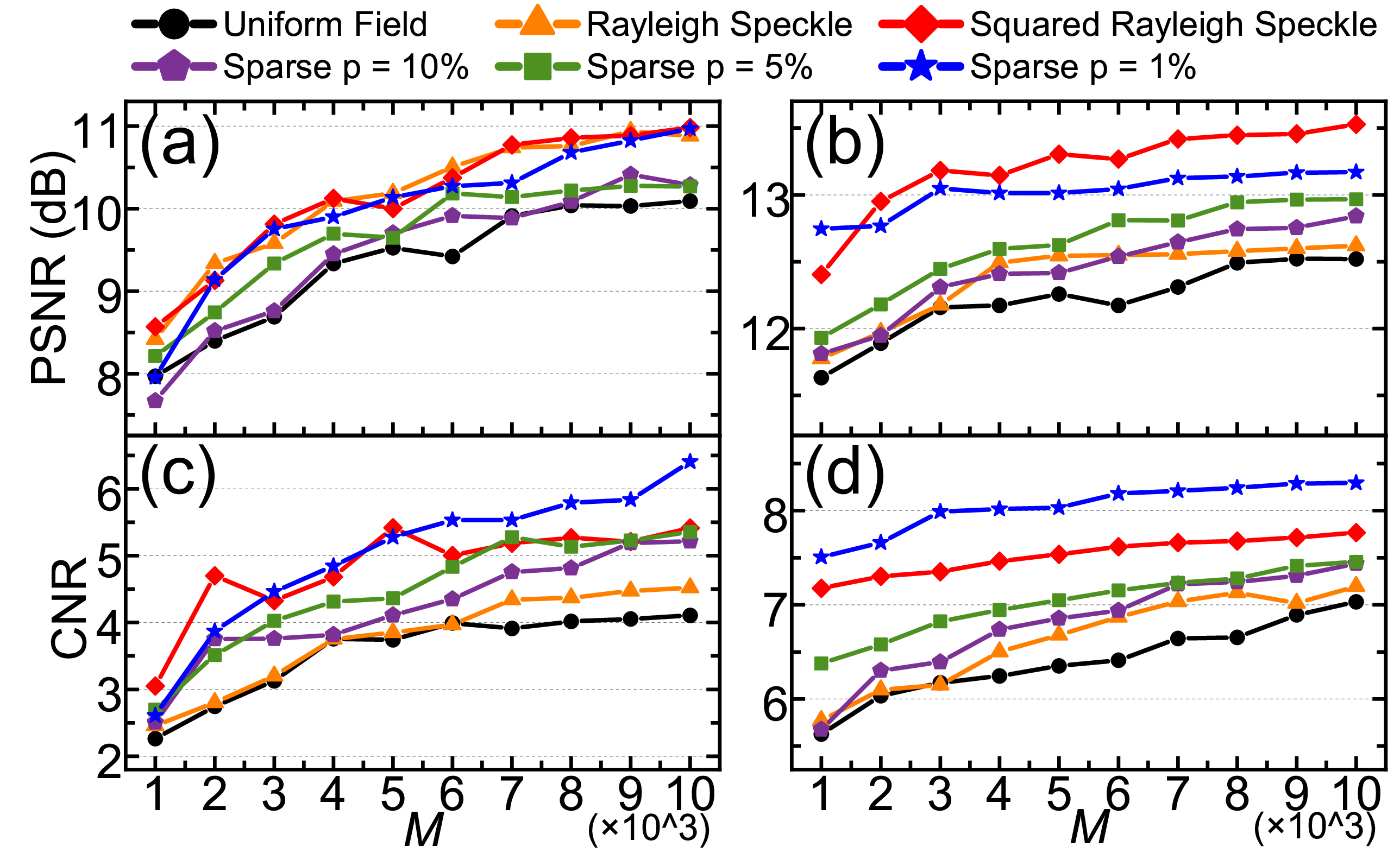}
	\caption{PSNR and CNR of the reconstructed binary object [Fig.~\ref{fig01:setup}(b)] as functions of the number of measurements $M$. Panels (a,b) show the PSNR, and panels (c,d) show the CNR, obtained using intensity-correlation reconstruction (a,c) and nonnegative TV-regularized reconstruction (b,d). Results are shown for six target fields: a uniform field, Rayleigh speckle, squared-Rayleigh speckle, and sparse binary fields with $p=10\%$, $5\%$, and $1\%$. Different y-axis ranges are used for the two reconstruction methods to resolve their respective trends.}\label{fig06:Npsnr}
\end{figure}

\begin{figure}[t]
	\centering
	\includegraphics[width=0.95\columnwidth]{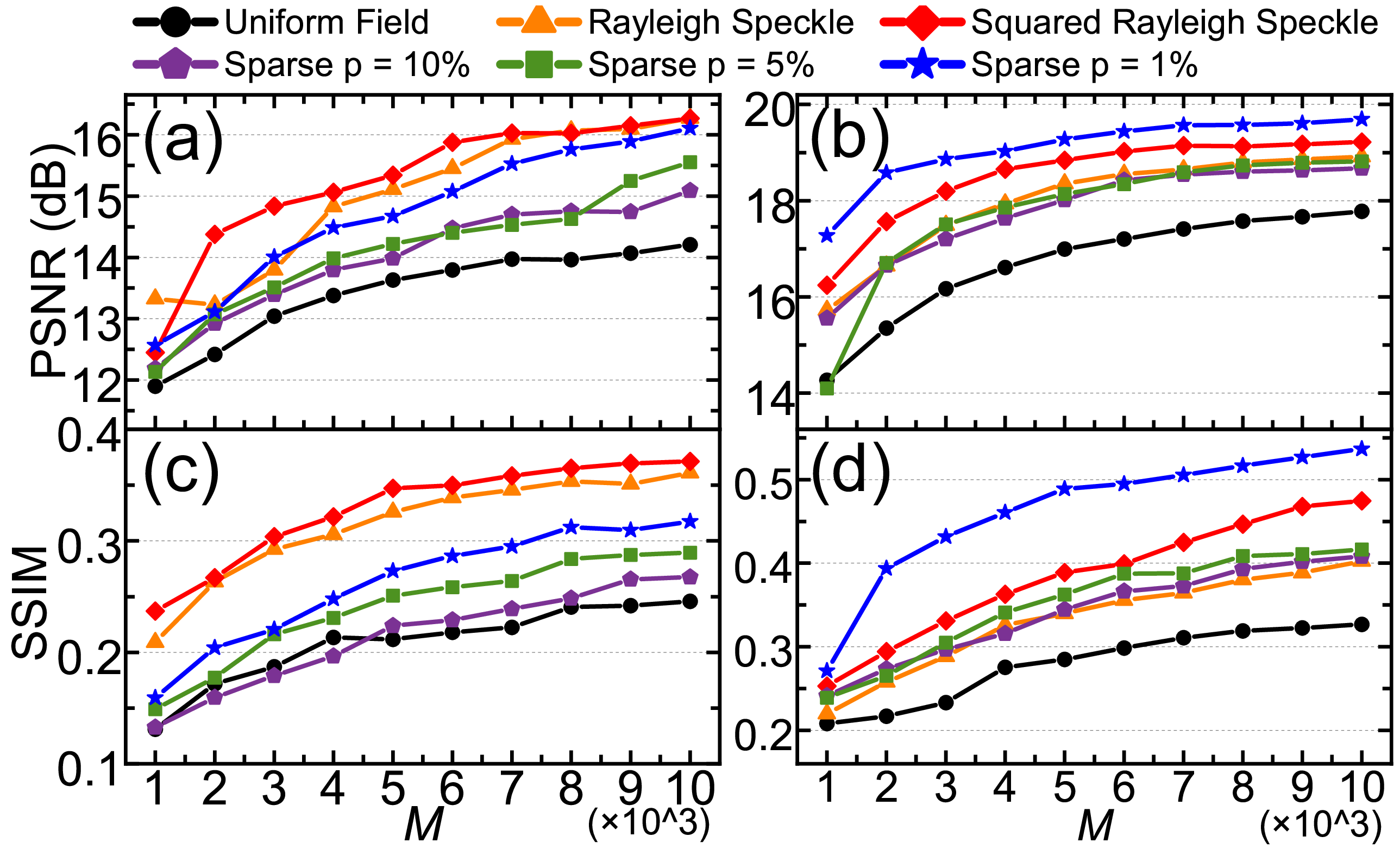}
	\caption{PSNR and SSIM of the reconstructed grayscale object [Fig.~\ref{fig01:setup}(c)] as functions of the number of measurements $M$. Panels (a,b) show the PSNR, and panels (c,d) show the SSIM, obtained using intensity-correlation reconstruction (a,c) and nonnegative TV-regularized reconstruction (b,d). Results are shown for the same six target fields as in Fig.~\ref{fig06:Npsnr}, with different y-axis ranges used for the two reconstruction methods to resolve their respective trends.}\label{fig07:Gpsnr}
\end{figure}

To quantify the effects of target-field design in dynamic multi-aperture SPI, we increased the number of measurements to $M=10{,}000$. Figure~\ref{fig06:Npsnr} shows the PSNR and CNR for the binary object, whereas Fig.~\ref{fig07:Gpsnr} shows the PSNR and SSIM for the grayscale object, all as functions of $M$. The same six target-field configurations were used for both objects.

For both reconstruction methods in Fig.~\ref{fig06:Npsnr}, the PSNR and CNR generally increased with $M$, despite local fluctuations, with the rate of improvement tending to decrease at larger $M$. Nonnegative TV regularization produced higher PSNR and CNR values than intensity-correlation reconstruction across the tested target fields and measurement range. For intensity-correlation reconstruction [Figs.~\ref{fig06:Npsnr}(a) and~\ref{fig06:Npsnr}(c)], the leading PSNR alternated among Rayleigh speckle, squared-Rayleigh speckle, and the $p=1\%$ sparse field. In contrast, the $p=1\%$ sparse field generally yielded the highest CNR for moderate and large values of $M$. Under nonnegative TV-regularized reconstruction [Figs.~\ref{fig06:Npsnr}(b) and~\ref{fig06:Npsnr}(d)], squared-Rayleigh speckle produced the highest PSNR for most values of $M$, whereas the $p=1\%$ sparse field consistently achieved the highest CNR. These results show that high-contrast target fields generally improved reconstruction quality, but the optimal target-field statistics depended on the evaluation metric.

For the grayscale object, both PSNR and SSIM generally increased with $M$ for either reconstruction method as shown in Fig.~\ref{fig07:Gpsnr}, although the improvements became more gradual at larger $M$. TV regularization increased PSNR throughout the tested range and generally improved SSIM. Under intensity-correlation reconstruction [Figs.~\ref{fig07:Gpsnr}(a) and~\ref{fig07:Gpsnr}(c)], squared-Rayleigh speckle generally yielded the highest PSNR and SSIM, while the $p=1\%$ field performed best among the sparse binary fields. The continuous-valued target fields showed only modest additional improvements at large $M$, whereas some sparse-field curves continued to increase. Under nonnegative TV-regularized reconstruction [Figs.~\ref{fig07:Gpsnr}(b) and~\ref{fig07:Gpsnr}(d)], the $p=1\%$ sparse field consistently achieved the highest PSNR and SSIM, followed by squared-Rayleigh speckle. These results demonstrate that the target-field configuration providing the highest grayscale-image quality depended strongly on the reconstruction algorithm. These results support task-specific design of the coded illumination.

Reconstruction quality depends on the measurement budget, target-field statistics, and reconstruction algorithm. Intensity contrast and spatial correlations may affect modulation depth and the conditioning of the sensing operator, while TV regularization introduces an object-gradient prior. These contributions were not isolated experimentally. Nevertheless, the algorithm- and metric-dependent rankings show why illumination and reconstruction should be designed jointly. Here, numerically propagated DMD patterns served as computational references; directly using the prescribed target fields remains to be evaluated. Joint optimization of the aperture distribution, reference patterns, and reconstruction algorithm may improve fidelity and measurement efficiency.

The broader significance is the integration of dynamic coded-aperture illumination and single-pixel detection in a lensless acquisition system. The DMD--detector--acquisition realizes the binary-modulation route and enables high-speed image acquisition. Coherent propagation converts binary aperture modulation into programmable grayscale illumination without relay projection optics or temporal grayscale synthesis. This separates spatial encoding from detector-array fabrication and allows the illumination and detector to be selected for the imaging task. The architecture is potentially useful where focal-plane arrays are costly or inefficient, or relay projection optics are impractical. Extension beyond the demonstrated wavelength of 632.8 nm would require compatible sources, modulator response, detector sensitivity, and hologram design.

In summary, we experimentally realized lensless SPI using dynamic coded-aperture illumination and a single-pixel detector. High-frame-rate binary DMD masks synthesize grayscale fields without relay projection optics. Experiments with six target fields and two reconstruction methods establish the importance of matching aperture encoding to the object and computational recovery. This architecture combines rapid binary modulation and flexible grayscale illumination with single-pixel readout, providing a basis for adapting lensless imaging to different detection requirements.

The supplementary material provides metric definitions of PSNR, CNR and SSIM values. It also details binary-amplitude hologram generation and diffraction at \(p=1\%\) sparsity, including a representative target field, its binary hologram, and the corresponding Huygens–Fresnel numerical diffraction simulation.

This work was supported by the National Natural Science Foundation of China (62105188).

\section*{AUTHOR DECLARATIONS}
\subsection*{Conflict of Interest}
The authors have no conflicts to disclose.

\subsection*{Author Contributions}
\textbf{Yuhan Guo}: Methodology (equal); Software (equal); Formal analysis (equal); Investigation (equal); Data curation (equal); Validation (equal); Visualization (equal); Writing -- original draft (equal).
\textbf{Shilei Li}: Methodology (equal); Software (equal); Formal analysis (equal); Investigation (equal); Resources (supporting); Data curation (equal); Validation (equal); Visualization (equal); Supervision (supporting); Funding acquisition (supporting).
\textbf{Tong Zhou}: Resources (supporting).
\textbf{Liming Li}: Conceptualization (lead); Methodology (supporting); Supervision (lead); Project administration (lead); Funding acquisition (lead); Writing -- original draft (equal); Writing -- review \& editing (lead).

\section*{DATA AVAILABILITY}
The data that support the findings of this study are available from the corresponding author upon reasonable request.

\section*{REFERENCES}
\bibliography{APL_references}
\end{document}